\documentstyle[prl,aps,epsf]{revtex}

\begin{document}

\title{Intertube coupling in ropes of single-wall carbon nanotubes}
\author{H.~Stahl$^*$, J.~Appenzeller$^*$, R.~Martel$^\dagger$, Ph.~Avouris$^\dagger$, B.~Lengeler$^*$}
\address{$^*$ 2.~Physikalisches Institut, RWTH Aachen, D-52056 Aachen, Germany \\ $^\dagger$ IBM T.~J.~Watson Research Center, Yorktown Heights, NY 10598, USA}
\date{\today}
\maketitle

\begin{abstract}
We investigate the coupling between individual tubes in a rope of single-wall carbon nanotubes using four probe resistance measurements. By introducing defects through the controlled sputtering of the rope we generate a strong non-monotonic temperature dependence of the four terminal resistance. This behavior reflects the interplay between localization in the intentionally damaged tubes and coupling to undamaged tubes in the same rope. Using a simple model we obtain the coherence length and the coupling resistance. The coupling mechanism is argued to involve direct tunneling between tubes.
\end{abstract}


The unique structural and electronic properties of carbon nanotubes make them interesting objects for basic science study as well as applications. The relation between their geometry and and electronic structure is of particular interest. Semiconducting or metallic behavior is possible depending on tube diameter and chirality \cite{Saito92}. Based on their unique properties, several applications in electronics have been proposed and some, such as field effect transistors \cite{Tans98b,Martel98} and diodes \cite{Yao99b} have already been demonstrated. While the electronic structure of individual tubes has been characterized using scanning tunneling spectroscopy and found to be in agreement with the theoretical predictions \cite{Wildoeer98}, the interaction between tubes in ropes has received much less attention. Some studies have concluded that the coupling between tubes must be weak \cite{Bockrath97}, but few attempted to directly measure this interaction \cite{Yao99b,Fuhrer00}. Thus, most of the applications rely on single tubes bridging metal contacts \cite{Tans98b,Tans97}. However, the extensive use of nanotubes in future nano-electronics would also require a knowledge of the tube-tube electronic coupling. 

\begin{figure}[ht]
\epsfxsize=7.5cm
\epsfbox{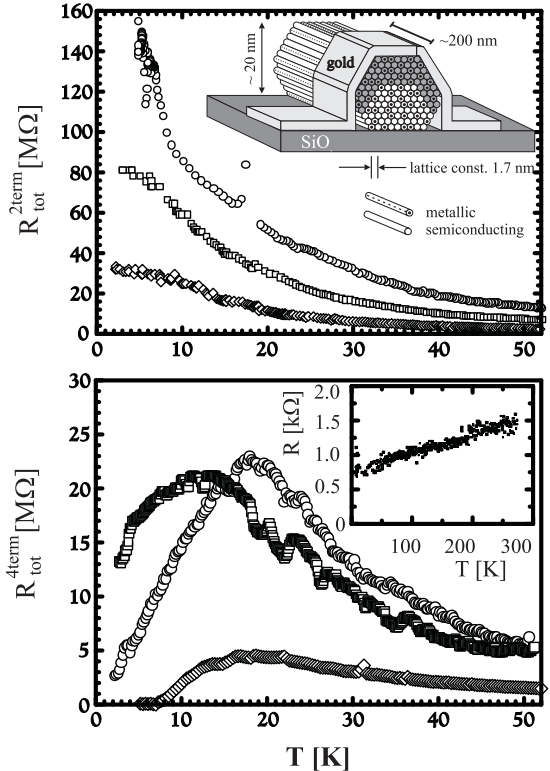}
\caption{\label{4term}Two (2t) and four (4t) terminal resistance versus temperature for three different samples. The insets show a schematic of a section of a rope and a 4t measurement on an undamaged rope. Note the very different values of the resistance in the main figure and the second inset.}
\end{figure}

Here, we present a novel approach that allows us to determine the electrical coupling between tubes in a rope using four terminal transport measurements. The ropes are self-assembled bundles of carbon nanotubes, in which the tubes line up parallel to each other. The tubes in our ropes have diameters close to 1.4 nm and form a regular triangular lattice with a lattice constanct of $d_0$ = 1.7 nm \cite{Thess96}. Both, semiconducting and metallic tubes are present in a rope in a random distribution. In our experiment, the ropes are dispersed on an oxidized Si substrate and gold electrodes were subsequently fabricated on top of the ropes (inset Fig.~\ref{4term}). The key feature in our investigation involves a sputtering of the rope before deposition of the electrodes by an Ar$^+$ ion beam at an energy of 500 eV. The purpose of the sputtering is to introduce defects into the top nanometers of the rope. As will be shown later, this will enable us to vary the path taken by the electric current in a well defined manner. In order to estimate the extent of the sputter damage, a Monte Carlo simulation was performed \cite{TRIM}. From our sputtering conditions, we estimate that the damage reaches about 6 ($\pm$2) nm deep into the rope and the damage density is about one defect per 1000 atoms, which gives a distance of 5-10 nm between defects along the tubes. This defect density is high enough to have a significant influence on the electrical properties of the tubes, while at the same time, it is low enough to preserve their structural integrity. The damage in the upper part of the rope is confined to the area directly underneath the gold contacts (note the shading in the inset of Fig.~\ref{4term}), while the main part of the rope between the electrodes was not exposed to the ion beam and is thus undamaged. Electronic transport in the damaged metallic tubes is strongly affected by the defects, while contributions from semiconducting tubes are negligible at the low temperatures used in the experiments (the typical band gap fo
r
 semiconducting tubes of $\sim$ 1.4 nm diameter is $\sim$ 500 meV \cite{Saito92,Wildoeer98}).

The resistances were measured using standard lock-in techniques while the sample was cooled in a $^4$He continuous flow cryostat with a base temperature of 1.5 K. We fully characterized a total of 13 samples. Typical results of R vs.~T curves are presented in Fig.~\ref{4term}. The two terminal (2t) resistances were found to increase with decreasing temperature. On the other hand, the four terminal (4t) measurements showed a pronounced resistance maximum at temperatures around 20 K in all of the samples we made. This behavior is caused by the damage introduced by sputtering. As a comparison, a 4t measurement of an undamaged rope is shown in the second inset in Fig.~\ref{4term}. In this case, we observe a decrease in resistance with decreasing temperature over the whole temperature range. Thus, the undamaged ropes show a metallic behavior as is expected for ropes consisting, at least in part, of metallic tubes \cite{Kane98}. The linear dependence of the resistance on the temperature is attributed to phonon scattering \cite{Kane98}. It is important to note the very different values of the resistance in the damaged and undamaged ropes, the latter being only of the order of 1 k$\Omega$, while the former is in the range of several M$\Omega$. Thus, the damage greatly increases the resistance, but it does not block the electrical transport. Obviously, the damaged, metallic tubes under the gold contacts carry most of the current, since only a few k$\Omega$ is expected for undamaged nanotubes in the rope, and the semiconducting ones, damaged or undamaged, are insulating. The metallic tubes are one-dimensional systems with two degenerate modes at the Fermi energy \cite{Saito92}, hence the resistance of a segment of length $L$ containing defects with an average distance $L_0$ is given by $R = h/4e^2\,\cdot\,L/L_0$ (neglecting any interference effects). The extent of the damaged areas along the direction of current transport, i.~e.~the width of the gold electrodes, is typically 200~nm. The rope segments between the contacts 
a
re undamaged and thus their contribution to the resistance is negligible. At room temperature, we find resistances of the ropes around 200~k$\Omega$. This resistance corresponds to a mean free path of $L_0 \simeq 6\ nm$, which is contistent with the mean distance between damaged sites obtained by the Monte carlo simulation.

In Fig.~\ref{4term}, both 2t and 4t measurements show an increase in resistance when cooling the sample. This increase is caused by electronic localization in the damaged tube, an interference effect which increases the resistance by coherent backscattering of electrons by the defects. Localization occurs when the phase coherence length, $L_\Phi$, exceeds the average distance between scatterers $L_0$ in the sample. There is strong localization if $L_\Phi$ exceeds the localization length $L_C = M\cdot L_0$, with M being the number of modes in the system \cite{Datta}. In this case all modes are strongly affected by the localization and the resistance increases exponentially with decreasing temperature. In view of the short mean free path $L_0 \simeq 5-10~nm$ and M=2, the rapid increase in resistance at low temperatures is likely to be caused by a strong localization in the damaged metallic nanotubes in the rope.

Below a sample-dependent temperature around 20 K, the resistance obtained by the 4t measurement starts to decrease. This effect cannot be caused by some gold-tube contact resistances, since these do not contribute in a 4t measurement. Any scattering mechanism (e.~g.~phonon scattering), which could possibly lead to such a behaviour, would show in both, the 4t and the 2t measurement. In general, the absence of a similar decrease in the 2t measurement proves, that this behavior can not be caused by a change in transport inside the actual current path, i.~e.~the damaged tube. We will now discuss a model how to understand our experiments and will extract information about tube-tube interactions.

Key to the understanding of our experiments is the realization that disorder can switch the current path from one tube to another inside the rope. Normally, we expect the current to be carried by the tube that has the lowest contact resistances to the electrodes. This must be a tube at the surface of the rope in direct contact with the electrodes. In our experiments, the surface tubes (and all other tubes about 6 nm deep into the rope) were damaged during the sputter treatment and thus show high resistance already at room temperature. When the sample is cooled the resistance increases due to localization in the damaged tube, and eventually grows sufficiently high, so that the current switches its path to another, undamaged metallic tube deeper inside the rope. The tubes are only weakly coupled, so the coupling resistances are rather high and it is for this reason that the undamaged tubes do not carry the current at 300 K. Once, however, the resistance in the damaged surface tube grows higher than the coupling resistance, the current favors this 'new' path and switches to the undamaged tube in the bulk of the rope. In a 4t measurement, we will then only detect the transport inside the bulk tube, which involves a resistance of a few k$\Omega$, while in the 2t measurement both channels are highly resistive, one due to localization, and the other due to the high coupling resistances involved in changing the current path.

\begin{figure}[ht]
\epsfxsize=7.5cm
\epsfbox{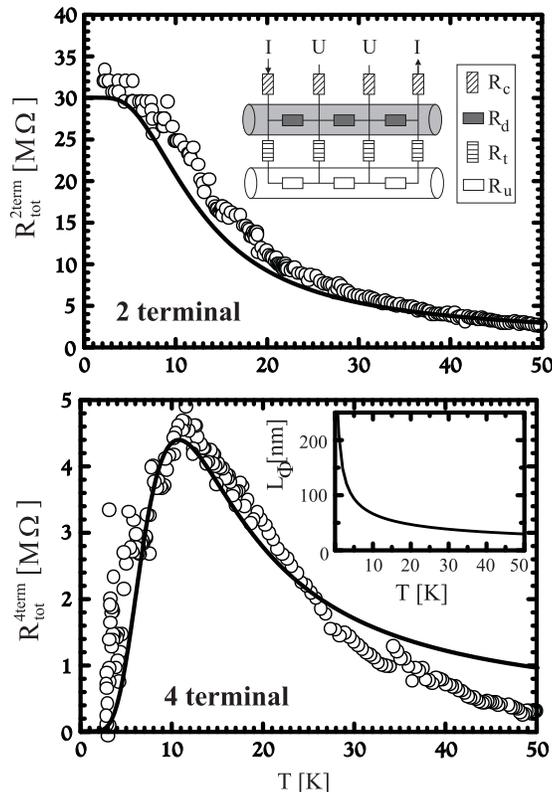}
\caption{\label{simulation}Comparison between measurement (symbols) and fit (solid line) according to the model described in the text. The insets show the resistance network of the model and the temperature dependence of the phase coherence length used in the fits.}
\end{figure}

We will now develop our model in order to be able to analyze the observations quantitatively. Consider a network of damaged and undamaged tubes with resistances $R_d$ and $R_u$ with the inter-tube coupling resistance $R_t$ and with contact resistances $R_c$ which connect the damaged tube at the surface to the gold electrodes. The inset in Fig.~\ref{simulation} shows how these resistances are connected in the 4t model. In order to calculate the total 2t and 4t resistances, we have to evaluate the individual resistances. $R_d$ is governed by strong localization and can therefore be described by \cite{Datta} \begin{displaymath}R_d = \frac{L}{L_\Phi} \cdot \frac{h}{2e^2} \cdot \frac{1}{2} \left(e^{\frac{2L_\Phi}{M\cdot L_0}}-1\right).\end{displaymath} $L_\Phi$ follows a power law dependence on temperature, $L_\Phi \sim T^{-\alpha}$, and we will use this to describe $R_d$. The inter-tube coupling resistance $R_t$ will be taken to be independent of temperature (we will justify this later on). The coupling resistances are placed underneath the electrodes, i.~e.~connected to the damaged areas of the surface tube, since a change in current path only occurs, where transport inside the surface tube gets 'blocked' by localization. The resistance of the undamaged tube $R_u$ is much smaller than $R_t$ and $R_d$ (cf. the second inset in Fig.~\ref{4term}) and since it is always connected to $R_t$, it does not play any significant role. The last resistance to be discussed is the contact resistance $R_c$ between the gold electrode and the surface tube. This resistance will only show up in 2t measurements but will not contribute to the 4t resistance. We will neglect this resistance for the moment and we will justify this below.

Figure \ref{simulation} shows experimental data from one of our samples, together with the fits based on our model. The results of both 4t and 2t measurements are well reproduced, underlining the validity of our simple model. First, we note that indeed no additional contact resistances $R_c$ are necessary to describe the 2t measurements. Second, from the fits in Fig.~\ref{simulation} we can extract $L_\Phi(T)$ as a function of temperature as shown in the inset. The coherence length turns out to be about 200 nm at the lowest temperature, a value significantly lower than that reported by other groups \cite{Bockrath97,Tans97,Shea00}, but the discrepancy is not surprising. We are dealing here with a disordered system. It is well known that disorder significantly enhances phase breaking processes \cite{Altshuler82}. We find that the temperature dependence of $L_\Phi$ can be described by $L_\Phi \sim T^{-\alpha}$ with $\alpha = 0.33 - 0.5$. $\alpha = 1/3$ points to dephasing by disorder-enhanced electron-electron scattering \cite{Altshuler82}, while $\alpha = 1/2$ suggests electron-phonon scattering. Both processes seem to be involved, with the electron-electron scattering possibly becoming dominant at the lowest temperatures \cite{Shea00}.

Next we evaluate the coupling resistance $R_t$ between the tubes. $R_t$ is extracted from the data in a very simple manner and turns out to be the most reliable and stable parameter in the simulation, since it is only determined by the {\em value} of the resistance maximum in the 4t measurement. The temperature dependence  of $R_d$ determines the {\em shape} and {\em position} of the maximum in temperature ($R_d(T_{max}) \simeq R_t$). Since slight variations in sputter damage ($L_0$) significantly affect $R_d$ there is no strict correlation between $R_t$ and $T_{max}$, but in spite of this $R_t$ can be obtained from the value of the resistance maximum. Analyzing $R_t$ for our 13 samples, we obtained values ranging from 2 M$\Omega$ to 140 M$\Omega$. Which coupling mechanism can explain such an enormous range of values? Hopping processes are sometimes invoked to describe inter-tube transport \cite{Kaiser99}. This involves transport by hopping through intermediate states (e.~g.~via other tubes). In our case, a single transfer would correspond to a resistance of about 2 M$\Omega$, and thus the highest value of 140 M$\Omega$ would need 70 transfers, barely imaginable with only 100 tubes in the rope at all. Moreover, the hopping processes are thermally activated and thus the coupling between tubes would eventually freeze out, leaving the rope insulating at the lowest temperatures, in contrast to the observation. So, the only explanation that fits our data seems to be a tunneling process between the tubes. In this process a small range of distances between the tubes leads to a large range of resistances due to the exponential dependence typical for tunneling. Furthermore, this process does not freeze out even at the lowest temperatures.

We will now try to link the experimental findings for $R_t$ to the geometry of the rope, i.~e.~the distances between the bulk and the surface tubes. The cross section of a rope consists of typically 100 single tubes. A fraction of about 2/3 of the tubes is semiconducting, while the remaining 1/3 is said to be metallic \cite{Saito92}. The tubes in the top part of the rope are damaged in the sputter treatment. When the resistance in the damaged surface tube has increased sufficiently (by localization) the current can switch via the coupling resistance into an undamaged metallic tube in the bulk, which involves tunneling over some distance d within the triangular lattice of the rope. Of course, the depth of the damage of the sputter treatment (about 6 nm) sets a lower limit for the distance d within which we can find an undamaged metallic tube. How can we describe the coupling resistance $R_t$ that is caused by the tunnel process ? For the coupling of one dimensional wave guides separated by a tunnel barrier we find \cite{Eugster91} $R_t = h/4e^2 \cdot v_F/v_\perp \cdot 1/T \simeq h/4e^2 \cdot e^{2 \kappa d}$. The velocity perpendicular to the tube axis $v_\perp$ can be approximated by the Fermi velocity $v_F$ when transport along the damaged tube is blocked by localization. The transmission T in the tunnel process is determined by the overlap of the wave functions of the tubes, with $\kappa$ being related to the barrier height. Given the linear dispersion relation for the metallic nanotubes $\epsilon (k) = \hbar v_F (k-k_F)$ around the Fermi energy and the barrier height $\Phi$, $\kappa$ is calculated as $\kappa=\Phi/(\hbar\cdot v_F)$. Since the electrons tunnel through the other tubes, which are mostly semiconducting, the barrier height is given by the conduction band edge of these semiconducting tubes. All nanotubes share the same graphene structure, hence their work function is expected to be nearly the same \cite{Fuhrer00}, and the Fermi level of the metallic tubes is expected to align midgap the semiconductin
g
 energy gap. with an average band gap of 500 meV we find $\Phi = E_{gap}/2 \simeq 250 meV$. Using $v_F = 10^6\ m/s$ \cite{Schoenenberger99}, we obtain a penetration depth of $1/2\kappa = 1.25\ nm$, comparable to the value given in reference \cite{Maarouf00}.

\begin{figure}[ht]
\epsfxsize=7.5cm
\epsfbox{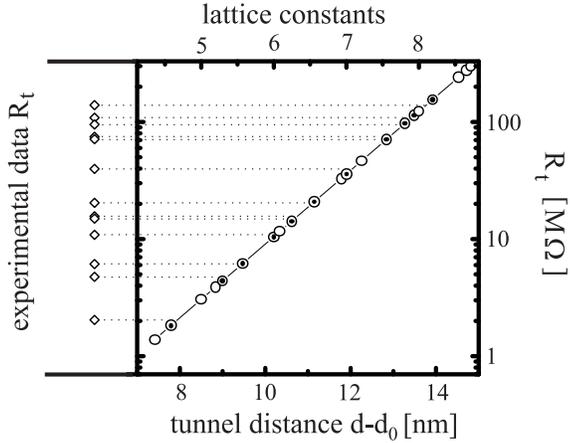}
\caption{\label{tunneling}Comparison between experimentally found coupling resistances $R_t$ (symbols on the left hand side) and values allowed by the theory (circles) for tunneling between tubes. Filled circles mark coincidences. $d_0$ = 1.7 nm is the lattice constant of the triangular lattice of the rope.}
\end{figure}

Figure \ref{tunneling} compares theoretical and experimental data, where the theoretical predictions result from evaluating the above formula for the discrete distances $d$ realized in the triangular lattice of the rope. (In fact, the tunnel distance is $d-d_0$ if d is the distance between the centers of the involved tubes. We used $\Phi = 225\,meV$. Since $k_BT \ll \Phi$, the tunnel resistance is indeed temperature independent.) We find that all data points coincide with values allowed by the model. None of the resistances we found corresponds to a distance shorter than about 8 nm, which is caused by the depth of the sputter damage of about 6 nm leaving no undamaged metallic tube in a shorter distance of the surface. Thus, the theoretical assumption of direct tunneling yields a consistent picture for the electronic coupling in nanotube ropes.

In conclusion, a sputter treatment of single-wall carbon nanotube ropes before making electrical contact resulted in damage and strong localization in the current carrying tubes at the surface of the rope. Below a sample specific threshold temperature the current tunnels into an undamaged, metallic tube in the bulk of the rope, leading to a dramatic reduction of the four terminal resistance. The value of the resistance maximum is related to the inter-tube coupling resistance between the involved tubes. Using a simple model, this inter-tube resistance is shown to be caused by direct tunneling between tubes.

\end{document}